\documentclass[useAMS,usenatbib]{mn2e} 
\usepackage{aas_macros}
\usepackage{graphics}
\usepackage[pdftex]{graphicx}
\usepackage{epstopdf}
\usepackage{epsfig}  
\usepackage{natbib} 
\usepackage{dingbat}
\usepackage{float}
\usepackage{amsmath}
\usepackage{times}
\usepackage[varg]{txfonts}
\usepackage{verbatim} 
\usepackage{multirow,bigdelim} 
\usepackage{color}
\usepackage{vmargin}
\setmarginsrb{1.2cm}{1cm}{1.2cm}{1cm}{1cm}{1cm}{1cm}{1cm}

  \makeatletter
    \renewcommand{\paragraph}{\@startsection{paragraph}{4}{\z@}%
      {-3.25ex\@plus -1ex \@minus -.2ex}%
      {1.5ex \@plus .2ex}%
      {\normalfont\small\centering}}
     
    \renewcommand{\subparagraph}{\@startsection{subparagraph}{5}{\z@}%
      {-3.25ex\@plus -1ex \@minus -.2ex}%
      {1.5ex \@plus .2ex}%
      {\normalfont\small\centering}}
    \makeatother

\newcommand{\kms}{{ km~s$^{-1}$}}
\newcommand{\hMpc}{{ \textit{h}$^{-1}$~Mpc}}

\title[Structures in the ZOA]{Predicting Structures in the Zone of Avoidance}

 \author[Sorce et al.]{{Jenny G. Sorce$^{1,2}$\thanks{E-mail:\text{jenny.sorce@astro.unistra.fr / jsorce@aip.de}},
Matthew Colless$^{3}$,
Ren\'ee C. Kraan-Korteweg$^4$,
Stefan Gottl\"{o}ber$^2$}\\
$^1$Universit\'e de Strasbourg, CNRS, Observatoire astronomique de 
Strasbourg, UMR 7550, F-67000 Strasbourg, France\\
$^2$Leibniz-Institut f\"{u}r Astrophysik, 14482 Potsdam, Germany\\
$^3$Research School of Astronomy and Astrophysics, Australian National University, Canberra, ACT 2611, Australia\\
$^4$Department of Astronomy, University of Cape Town, 7700 Rondebosch, South Africa}

\begin{document}
\date{}

\pagerange{\pageref{firstpage}--\pageref{lastpage}} \pubyear{2017}
\maketitle
\label{firstpage}


\begin{abstract}
  \indent The Zone of Avoidance (ZOA), whose emptiness is an artifact of our Galaxy dust, has been challenging observers as well as theorists
  for many years. Multiple attempts have been made on the observational side to map this region in order to better understand the
  local flows. On the theoretical side, however, this region is often simply statistically populated with structures but no real attempt
  has been made to confront theoretical and observed matter distributions. This paper takes a step forward using constrained
  realizations of the local Universe shown to be perfect substitutes of local Universe-like simulations for smoothed high density peak
  studies. Far from generating completely `random' structures in the ZOA, the reconstruction technique arranges matter according to the
  surrounding environment of this region. More precisely, the mean distributions of structures in a series of constrained and random
  realizations differ: while densities annihilate each other when averaging over 200 random realizations, structures persist when
  summing 200 constrained realizations. The probability distribution function of ZOA grid cells to be highly overdense is a Gaussian with a
  15\% mean in the random case, while that of the constrained case exhibits large tails. This implies that areas with the largest
  probabilities host most likely a structure. Comparisons between these predictions and observations, like those of the Puppis 3 cluster,
  show a remarkable agreement and allow us to assert the presence of the, recently highlighted by observations, Vela supercluster at about
 180~\hMpc, right behind the thickest dust layers of our Galaxy.
 
\end{abstract}


\section{Introduction}

Galactic dust obscures a large fraction of the sky, called the Zone of Avoidance \citep[ZOA,][]{1961gala.book.....S}. This region, apparently devoid of galaxies, used to be
avoided by observers because of the inherent difficulties, resulting from the strong dust extinction, in analyzing the galaxies there. However, to understand local flows and the Cosmic Microwave Background \citep[CMB,][]{1965ApJ...142..419P} dipole \citep[e.g.][]{1971Natur.231..516H,1976BAAS....8Q.351C,1977PhRvL..39..898S,2014A&A...571A..27P}, coverage of the whole sky is essential. Numerous studies in the past have indeed shown the relation between surrounding structures and the CMB dipole \citep{1991AIPC..222..421L,2006ApJ...645.1043K,2006MNRAS.368.1515E,2011ApJ...741...31B,2012MNRAS.427.1994G}. The lack of data in the ZOA (which covers about 10--15\% of the sky depending on the wavelength of observations) means that astronomers must rise to this challenge.

Many local large-scale structures are indeed bisected by the Galactic Plane. Two of them, Perseus-Pisces and the Great Attractor, actually lie across the ZOA on exactly opposite sides of the Local Group \citep[e.g.][]{1994AJ....108..921P,1994A&A...286...17S,1997AJ....113..905P}. Because they are partially obscured, determining the winner in this tug-of-war on the Local Group constitutes a daunting task \citep[see e.g.][]{1992AGAb....7..177S,1994ASPC...67...81S}. Similarly, identifying the connections between structures lying across the ZOA, if they exist,
is essential to measure the longest structures and verify their compatibility both in terms of size and abundance with predictions from the standard cosmological model or from the observations of fluctuations in the CMB \citep{2000A&ARv..10..211K,2011MNRAS.417.2938S}. At present, most CMB dipole estimates assume a homogeneously filled ZOA or an extrapolation of structures on either side of the ZOA, which prevents them from achieving high precision measurements. Thus there exists a real need to map the structures across the ZOA \citep[see][for a complete review on the subject]{2000A&ARv..10..211K}.

A great deal of effort has already been deployed to observe galaxies within the ZOA \citep[e.g.][for a non-extensive list with the last two references as complete reviews]{1994ASPC...67...99K, 2014MNRAS.443...41W,2016AJ....151...52S,2006MNRAS.369.1741D,2016MNRAS.460..923R, 2016MNRAS.457.2366S,2000A&ARv..10..211K,2005RvMA...18...48K} or at least to reduce its extent by using observations at less-affected wavelengths \citep[e.g.][]{2012AJ....144..133S, 2014MNRAS.444..527S,2014ApJ...792..129N}. A complementary approach is to use observations
surrounding the ZOA to reconstruct the missing information in it \citep{1994ASPC...67..185H,2006MNRAS.373...45E}. In this paper, we propose a method for
quantifying the probability of the presence of structures in different areas of the ZOA. 

The study is conducted within the framework of
constrained simulations of the local Universe  \citep[e.g.][]{2001ApJS..137....1B,2010arXiv1005.2687G}.
Unlike ordinary cosmological simulations, these simulations stem from initial conditions constrained by observational data to resemble the
local Universe. These observational data can be either radial peculiar velocities of galaxies \citep[]{2002ApJ...571..563K,2003ApJ...596...19K, 2014MNRAS.437.3586S} or redshift catalogs \citep[]{2010MNRAS.406.1007L,2013MNRAS.435.2065H}. The initial conditions are retrieved either backwards  \citep{1987ApJ...323L.103B,1991ApJ...380L...5H,1992ApJ...384..448H,1993ApJ...415L...5G,1996MNRAS.281...84V,1998ApJ...492..439B,2008MNRAS.383.1292L} or forwards \citep{2013MNRAS.429L..84K,2013MNRAS.435.2065H,2013MNRAS.432..894J, 2013ApJ...772...63W}. In the latter case, the initial density field is sampled from a probability distribution function consisting of a Gaussian prior and a likelihood. A complete overview of the different methods is given in \cite{2014ApJ...794...94W} and an extension of the backwards method to non-Gaussian fields is proposed in \citet{1995MNRAS.277..933S}.

Recently, \citet{2016MNRAS.455.2078S} released simulations constrained with the second radial peculiar velocity catalog \citep{2013AJ....146...86T} of the Cosmic Flows project \citep[e.g.][]{2012ApJ...749...78T,2012ApJ...749..174C} using a backwards method. Although estimating peculiar velocities constitutes an observational challenge, there is a double advantage in using them: first they are highly linear and correlated on larger scales than the density field. Consequently, the Zone of Avoidance can still be recovered when it becomes larger than the correlation length of the density field with a sufficiently sampled velocity field; second, peculiar velocities are excellent tracers of the underlying gravitational field as they account for both the baryonic and the dark matter. The backwards method permits producing a large number of realizations in a short amount of (computing) time which is paramount for such statistical studies. Running the simulations from the large set of initial conditions produced from the method is the most demanding of computer time.

Actually, as shown in Section~2, running such constrained simulations is not an absolute necessity when the goal is to study the smoothed high density peaks of the large-scale structure at redshift zero. There is thus no reason either to use forwards methods with high computational demands to get the reconstruction of the local Universe with the non-linearities. Basically, studying the linear reconstruction of the local Universe obtained with a backwards method\citep[e.g.][]{1995ApJ...449..446Z,1999ApJ...520..413Z}, or more precisely the constrained realizations of the latter \citep{1987ApJ...323L.103B,1992ApJ...384..448H,1996MNRAS.281...84V}, to actually have access to structures in the ZOA up to large distances as well as to their probability, is sufficient. Indeed, the reconstruction technique by definition goes to the null field (the overdensity $\delta$ equals 0) in the absence of data or in presence of very noisy data while, by extension, the constrained realizations technique statistically restores the missing structures. Namely, while the Wiener Filter reconstruction presents smoother and smoother structures up to reaching the null field with increasing distances from the center of the dataset, the constrained realizations compensate this smoothing by combining the constraints with a random field thus allowing us to fully access structures even beyond the volume sampled by the dataset. Detailed equations are given in Appendix A. The goal of this paper is to show that this restoration is not statistically random and depends on the surrounding structures up to large distances. Hence, it is possible to predict where the structures are more likely to be in the ZOA, as shown in Section~3. In Section~4, theoretical predictions are compared with observations, and the conclusions of this work are presented in Section~5.
 

\section{Methodology}

\subsection{Preparing Constrained Realizations} 

The scheme used by the CLUES project to build constrained initial conditions and run the simulations is described in detail in
\citet{2016MNRAS.455.2078S}. Since only some parts of the process are of interest here, the steps to produce realizations constrained by
observational radial peculiar velocity catalogs and a brief description of their purpose are the only aspects recapitulated here:
\begin{enumerate} 
\item grouping of galaxies in the radial peculiar velocity catalog to  remove non-linear virial motions
  \citep[e.g.][]{2015AJ....149...54T,2015AJ....149..171T} that would  cause problems when using the linear reconstruction method;
\item minimization of the biases \citep{2015MNRAS.450.2644S} in the  grouped radial peculiar velocity catalog; and
\item production of linear density and velocity fields constrained by  the grouped and bias-minimized observational peculiar velocity  catalog combined with a random realization to statistically restore the missing structures using the Constrained Realization technique \citep[CR,][]{1987ApJ...323L.103B,1991ApJ...380L...5H,1992ApJ...384..448H} .
\end{enumerate}

The Planck cosmology framework \citep[$\Omega_m$=0.307, $\Omega_\Lambda$=0.693, H$_0$=67.77, $\sigma_8$~=~0.829,][]{2014A&A...571A..16P} is used as a prior.

\subsection{Constrained Realizations versus Constrained Simulations}

Running constrained simulations is useful to study the formation and evolution of the large-scale structures as well as the small non-linear
scales. However, when interested only in the large-scale structure present today, there is no need to simulate the full formation history
of these structures---the linear reconstruction is sufficient. This reconstruction can be obtained with the Wiener Filter technique
\citep[linear minimum variance estimator, abbreviated as WF;][]{1995ApJ...449..446Z,1999ApJ...520..413Z}. Where data are
available, the WF-reconstructed field reproduces the observed large scale structure \citep[e.g.][]{1999ApJ...520..413Z,2012ApJ...744...43C}. The WF method
is also able to reconstruct across the ZOA provided that the latter does not cover too large a volume \citep[i.e. nearby, at low
redshift;][]{1994ASPC...67..185H,2000ASPC..218..173Z}. The further from the observer, the worse the extrapolation is, and in fact structures are
not reconstructed in the ZOA at large distances---the WF simply goes to the null field. By contrast, the CR technique restores statistically structures in that zone by adding a random field.

The goal of this paper is to determine if there are some regions in the ZOA that have a higher probability of hosting a structure than
others. To pursue this work, statistical analyses are required. This constitutes another argument in favor of studying the structures
directly in the fields obtained with the CR technique, since running lots of constrained and random (for statistical comparison) simulations
consumes a lot of computer time. One might argue that CR fields are not suitable substitutes for full simulated fields at the level of interest
for this paper, and of course CR fields present only the linear structures and obviously are in no way exact substitutes for the fully
simulated fields. However, as long as CR fields and simulated fields present a very similar large-scale structure for high-density peaks in
the ZOA, one can be used for the other in statistical tests.

\begin{figure}
\centering
\hspace{-1.54cm}\includegraphics[width=0.56\textwidth]{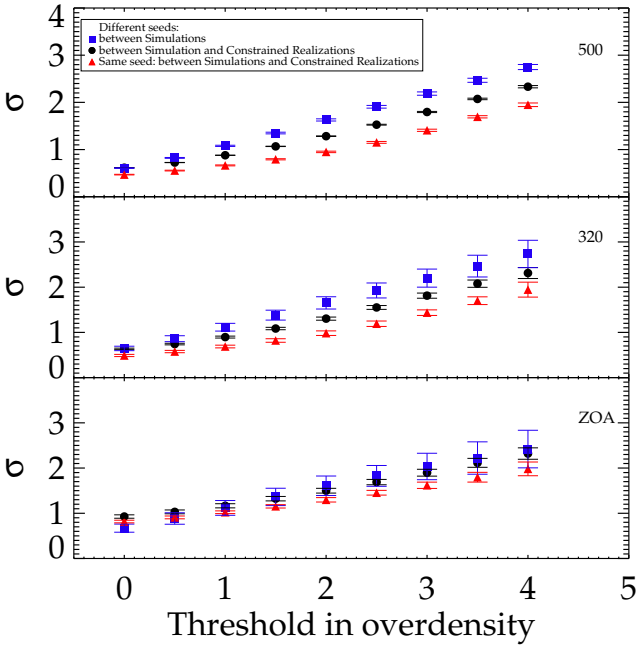} 
\vspace{-1cm}
\caption{Mean (points) and scatter (error bars) of the standard deviations ($\sigma$) obtained from cell-to-cell comparisons carried out on pairs of overdensity fields smoothed on a 5\hMpc\ scale at
  $z$=0 for the whole 500\hMpc\ box (top), for the 320\hMpc\ sub-box (middle), and for the Zone of Avoidance (bottom). The mean standard deviations are given as a function of the overdensity threshold. Pairs consist of two constrained simulations (filled blue square), one constrained simulation and one constrained realization (filled black
  circle) and one constrained simulation and one constrained realization sharing the same random part (filled red triangle). }
\label{fig:CRvsSimu}
\end{figure}

To demonstrate that CR fields can be taken as substitutes for simulated fields for our purpose, 15 simulations from \citet{2016MNRAS.455.2078S}
and the 15 corresponding CR fields (so that simulations and CRs share the same random part) are compared. We apply a cloud-in-cell scheme on a
256$^3$ grid to the particle distributions of the 15 simulations at $z$=0 with a subsequent Gaussian smoothing on a scale of 5\hMpc. The
resulting smoothed density field of any constrained simulation is then converted to the overdensity field. The CR technique indeed only
produces the overdensity field. Each simulated overdensity field is then compared to that of another constrained simulation, to that of a
constrained realization, and to that of the constrained realization sharing the same random field. For each pair of each category
(simulation/simulation; simulation/CR; simulation/CR sharing the same random field), we build an overdensity-overdensity plot (i.e.\ cells of
overdensity field \#1 versus cells of overdensity field \#2). If the two fields were identical, all points would lie on the 1:1 relation (cell
value in field \#1 is equal to cell value in field \#2). The difference between two fields is determined as the one-sigma, hereafter 1$\sigma$,
scatter (or standard deviation) around this 1:1 relation. We repeat this procedure for all the pairs of each category and then the mean and the
variance of the 1$\sigma$ scatters are derived.

In the top panel of Fig.~\ref{fig:CRvsSimu}, the result is three points (blue, black and red symbols for each of the
categories simulation/simulation, simulation/CR, simulation/CR sharing the same random field), with error bars, computed only for cells with positive overdensities.  
The procedure is repeated in smaller sub-boxes of size 320~\hMpc, where most of the observational constraints are, and in the ZOA (i.e.\ cells are
compared only in this area: the ZOA is taken to be a 20$^\circ$ cone with the apex at the center of the box and oriented in the same direction in the
realizations as in the observations.  The resulting points are shown in the middle and bottom panels of Fig.~\ref{fig:CRvsSimu}. We repeat the process,
changing the threshold to select higher and higher overdensity regions.

Figure~\ref{fig:CRvsSimu} shows that the 1$\sigma$ scatters increase with the threshold in accord with the expectation that, although the
large-scale structure is reproduced, high-density peaks might be shifted by a few megaparsecs from one realization to another, and from the
simulation to the CR. In addition, because simulations also probe the non-linear regime, and hence allow larger fluctuations, the 1$\sigma$
scatters are higher when comparing two simulations than when comparing a simulation and a CR. That is, the chance of getting a large contrast
between cells coming from two different simulations is higher than for cells coming from a simulation and a CR; and indeed in the latter,
linear-regime, case the fluctuations are reduced.

The main result from Fig.~\ref{fig:CRvsSimu} is that the differences between two simulations, or between a constrained realization and a
simulation built with different random fields, are larger on average than the difference between the simulation and the constrained
realization sharing the same seed (at least for the high density peaks of interest in this paper). Admittedly, the differences are more
pronounced when considering the full boxes rather than the smaller boxes or the ZOA. In the first case, it is expected because of the decrease in
cosmic variance when working with constrained simulations compared to working with random simulations \citep[e.g.][]{2016MNRAS.455.2078S}. The inner parts of the boxes are
more constrained (have more constraints) and thus they resemble more to the local Universe as well as to each other in the different
simulations (hence the smaller 1$\sigma$ scatters than for the full boxes). This is also expected in the second case because the ZOA is not as
well constrained as the other parts of the box given the scarcity of the data, especially at increasingly higher distances. The difficulty of reconstructing across the ZOA increases linearly with the distance from the observer. In addition, the error bars of the 1$\sigma$ scatter increase with the reduction of the number of cells compared between two realizations,
(i.e. with the reduction of the compared zones).

In that context, the scatters between the simulation and the CR sharing the same random field are even lower than between two
simulations, or between a CR and a simulation with different random fields. This suggests that for our purpose (a statistical study of high-overdensity peaks in the ZOA) 
one can be taken as a substitute for the other. Taking CR fields as substitutes for the simulated
fields to derive statistics on the large-scale structure in the ZOA, we produce 200 CR fields on 256$^3$ grids as well as 200 random fields
(sharing the same random part as the corresponding CR field) on the same grid size at a small computational cost.


\section{Probability of Structures}

\subsection{Random vs Constrained Probabilities}

\begin{figure*}
\centering
\includegraphics[width=1\textwidth]{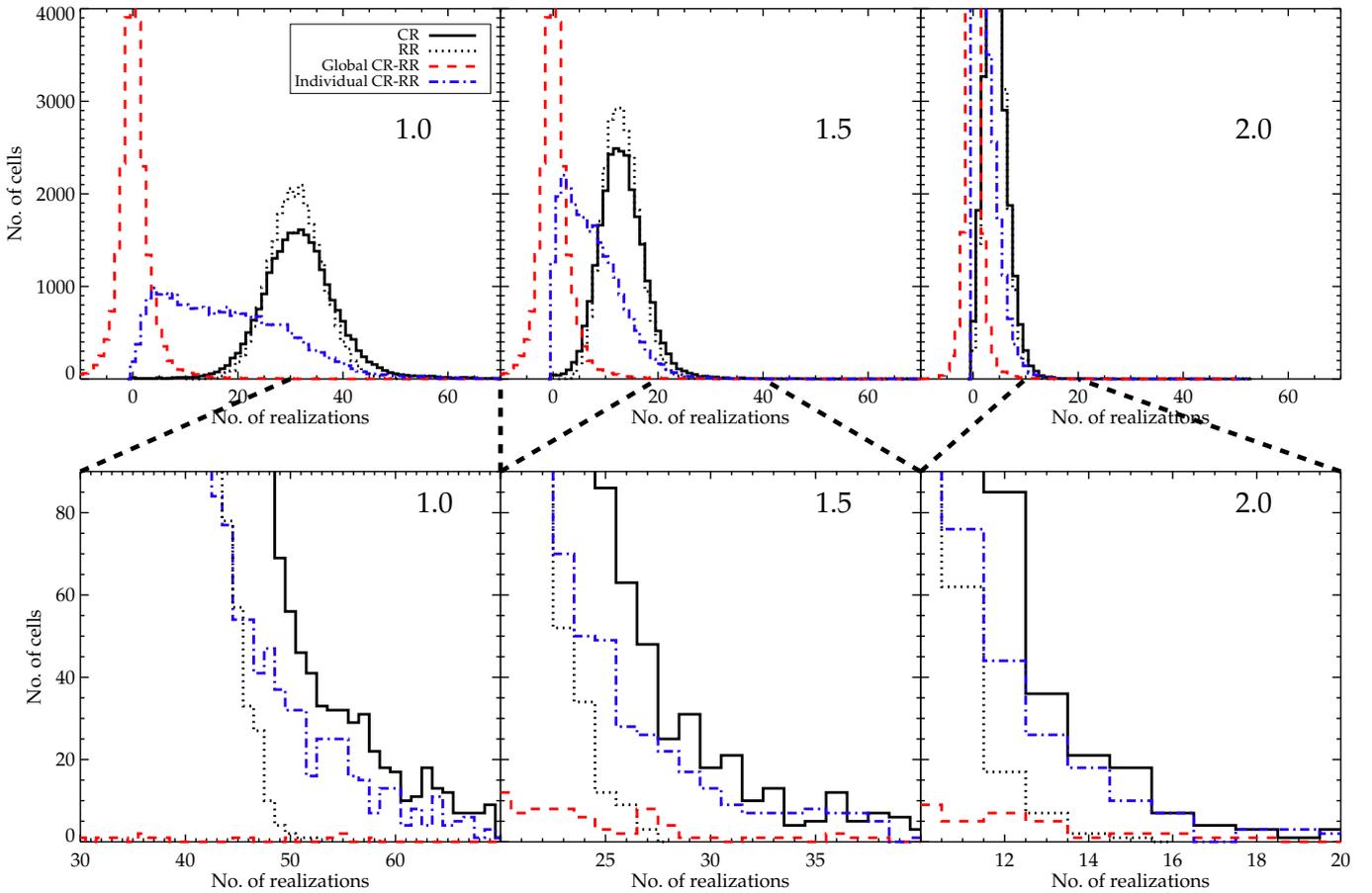}\\
\vspace{-.2cm}
\caption{Histograms of the number of cells with an overdensity above the threshold given in the top right corner of each panel (1.0, 1.5, or 2.0) as a function of the number of realizations. The dotted black line and solid black line represent, respectively, the probability distribution functions of cells in, respectively, random and
  constrained realisations having an overdensity value above the threshold. The bottom panels are zooms of the positive tails of the
  distributions. The red dashed line shows the distribution of the residual between the constrained and the random probabilities for a
  given cell. The blue dot-dashed lines are the distributions if cells are only counted if they exhibit an overdensity value above the
  threshold {\em only} in the CR and {\em not} in the RR sharing the same random seed. In this case a cell is counted only when the
  overdensity is entirely due to the constraints: the random part alone  would not have induced an overdensity above the given threshold. See
  the text for a more detailed explanation.}
\label{fig:hist}
\end{figure*}

The first goal is to quantify the constraining power of the CR technique in the ZOA. That is, at a given location in the ZOA, do the CR fields
present a structure with a greater probability than the random fields, or are the structures simply randomly distributed in a manner consistent
with the cosmological model?  With 200 constrained realizations and 200 random realizations, it is straightforward to study the probability that
a cell is overdense (i.e. hosting a structure) in both the random and constrained cases.

To determine these probabilities in the ZOA, cells belonging to the ZOA need first to be identified. Since we smooth the realizations over
5\hMpc, cells of 5\hMpc\ filling up the entire ZOA constitute the optimal choice. We split the ZOA into cells of 5\hMpc\ and compute their
overdensity value. We count the number of times a cell belonging to the ZOA has a value above a given threshold over 200 CRs. For clarity and
simplicity, the paraphrase `threshold-pass cell' is used to mean hereafter `a cell with a value above a given overdensity
threshold'. Finally, looping over all the ZOA cells, the number of cells that are threshold-pass in a certain number of CRs is obtained and the
probability distribution function of a cell to be threshold-pass is calculated. The same procedure is repeated with the random realizations
(RRs) to obtain the random probability distribution function of a cell to be threshold-pass.

If the constraints were not affecting the ZOA at all, i.e. if constraining the realizations was not giving some insight into the ZOA structures, then the constrained and random probability distribution
functions should be statistically identical. This would imply that the structures in the ZOA are completely randomly distributed and abide only by the prior cosmological model in both the random and the constrained cases. Figure~\ref{fig:hist} shows that this is not the case: the histograms of probabilities or the distributions of threshold-pass cells
(one threshold per panel in the top row) as a function of the number of realizations in which they are found differ between the constrained
(solid black line) and random (dotted black line) realizations. For instance, the top left panel of Fig.~\ref{fig:hist} shows that the
random probability for a cell to be threshold-pass at the 1.0 overdensity level follows a Gaussian distribution with a mean at 15\%
(30 out of 200 realizations, dotted black line). Thus there is on average about a 15$\pm$3\% chance of finding an overdensity greater than
1.0 in a cell. The constrained probability however does not follow a Gaussian distribution, the solid black line distribution is
skewed (skewness 1.5) and flatter (kurtosis 11), confirming that constraining the realization also affects the ZOA. The middle and right
panels of the top row on Fig. \ref{fig:hist} confirm that such observations are valid also for higher overdensity thresholds.

From these different probability distribution functions, it becomes clear that cells with probabilities in the high tail of the distribution
for constrained realizations, most likely, host a structure that is produced by the constraints surrounding the ZOA. The bottom row of
Fig.~\ref{fig:hist} zooms in on the high tail of the probability distribution functions calculated at a given threshold to highlight the existence of such cells.

To reinforce our claims, we pursue our quest for statistical differences between CRs and RRs in the ZOA by deriving two more probability distributions that are hereafter called the global probability and the individual probability:
\begin{itemize}
\item The global probability for a given cell is defined as the difference between its constrained and random probabilities. At a fixed overdensity threshold, the number of RRs in which a given cell is
  threshold-pass is subtracted from the number of CRs in which this cell is also threshold-pass. The procedure is repeated over all the different cells in the ZOA to derive the global probability
 distribution function.
\item The individual probability for a given cell is a bit more complex. Considering one CR, a given cell is counted as threshold-pass only in the case that in the RR sharing the same random seed as the CR, this same cell is {\em not} threshold-pass---i.e. the cell is threshold-pass in the CR only. In other words, we require that the constraints force the cell to become threshold-pass and the random component used in the CR technique to restore the missing structures  would not have induced an overdensity greater than the threshold in that cell. The process is then repeated over the 200 realizations and then over all the cells filling the ZOA to derive the individual probability distribution function.
\end{itemize}
 
The global and individual probability distribution functions are plotted in Fig.~\ref{fig:hist} as the red dashed and blue dot-dashed lines
respectively. These curves confirm the statistical difference between CRs and RRs in the ZOA. Whatever the threshold level is, the global
probability distribution function is a Gaussian centered on zero with (for instance) a standard deviation of 8\% for a 1.0 overdensity
threshold. This distribution implies that some constrained cells have on average a greater (or lower) probability to be threshold-pass than their random
counterparts. This indicates that structures (both over- and under-densities) are more likely to exist in the ZOA because of the constraints than can be explained as solely due to the prior
cosmological model and its associated statistical fluctuations. It confirms that the random and constrained probability distribution functions differ and that the probability of each cell
to be threshold-pass is different in the random and constrained cases. The individual probability distribution function goes further since the random seed is also fixed. With a mean
at 13\% and a standard deviation of 8\% for a 1.0 overdensity threshold, this distribution is even more remarkable: some cells are clearly
threshold-pass solely because of the constraints: the random realizations have a very low probability of generating overdensities in
the cells that are in the high tail of the individual probability distribution function on a 1-to-1 random seed basis.
 
Since there is a clearly demonstrated difference between threshold-pass cells in CRs and RRs, we can proceed to pursue the analysis of
structures in the constrained field to locate the most probable structures in the ZOA.
 
\subsection{Locations of Constrained Structures}

\begin{figure*}
\vspace{-0.9cm}
\centering
\hspace{-1.5cm} \includegraphics[width=0.55\textwidth]{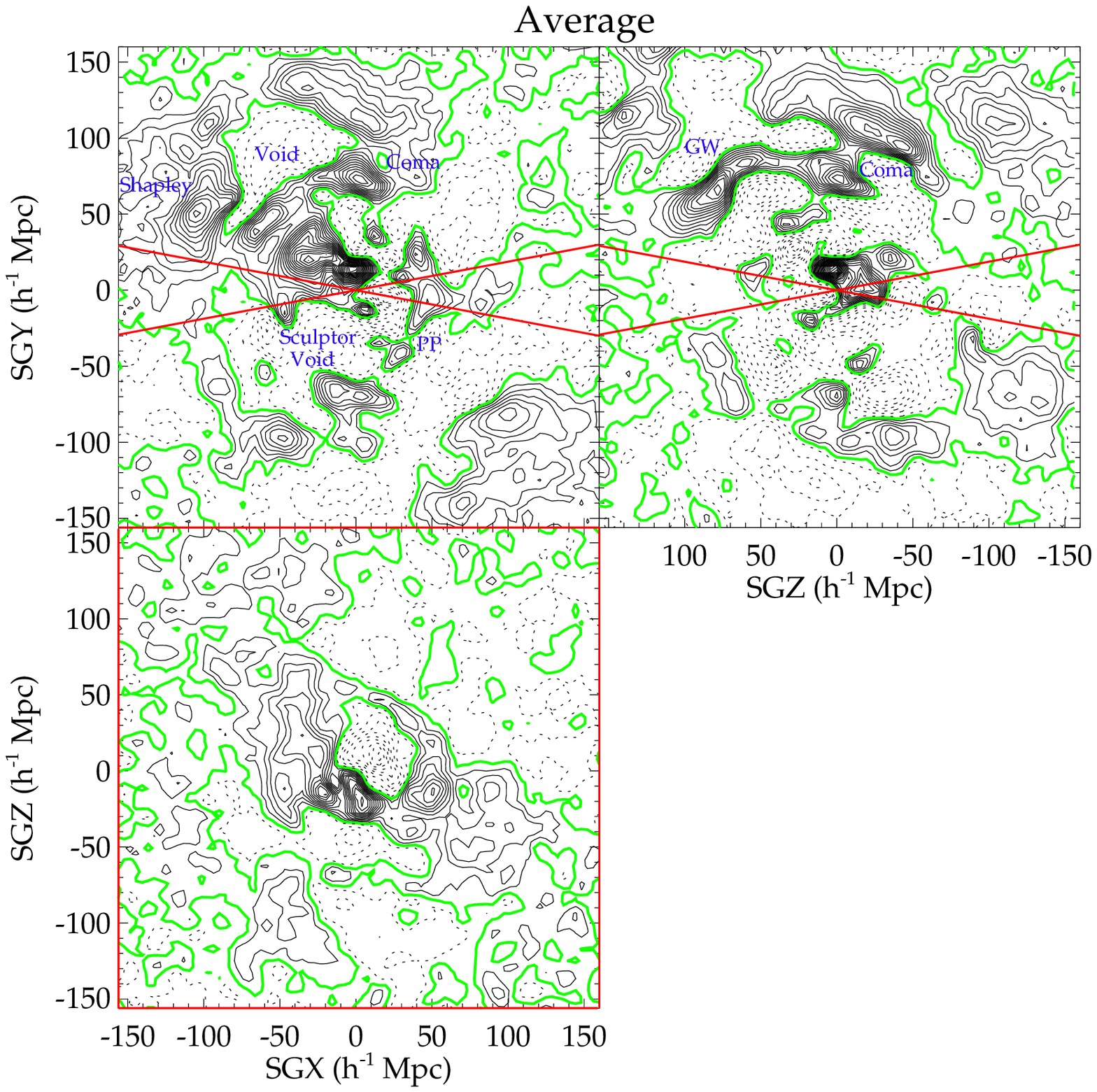}
\hspace{-0.58cm} 
\includegraphics[width=0.55\textwidth]{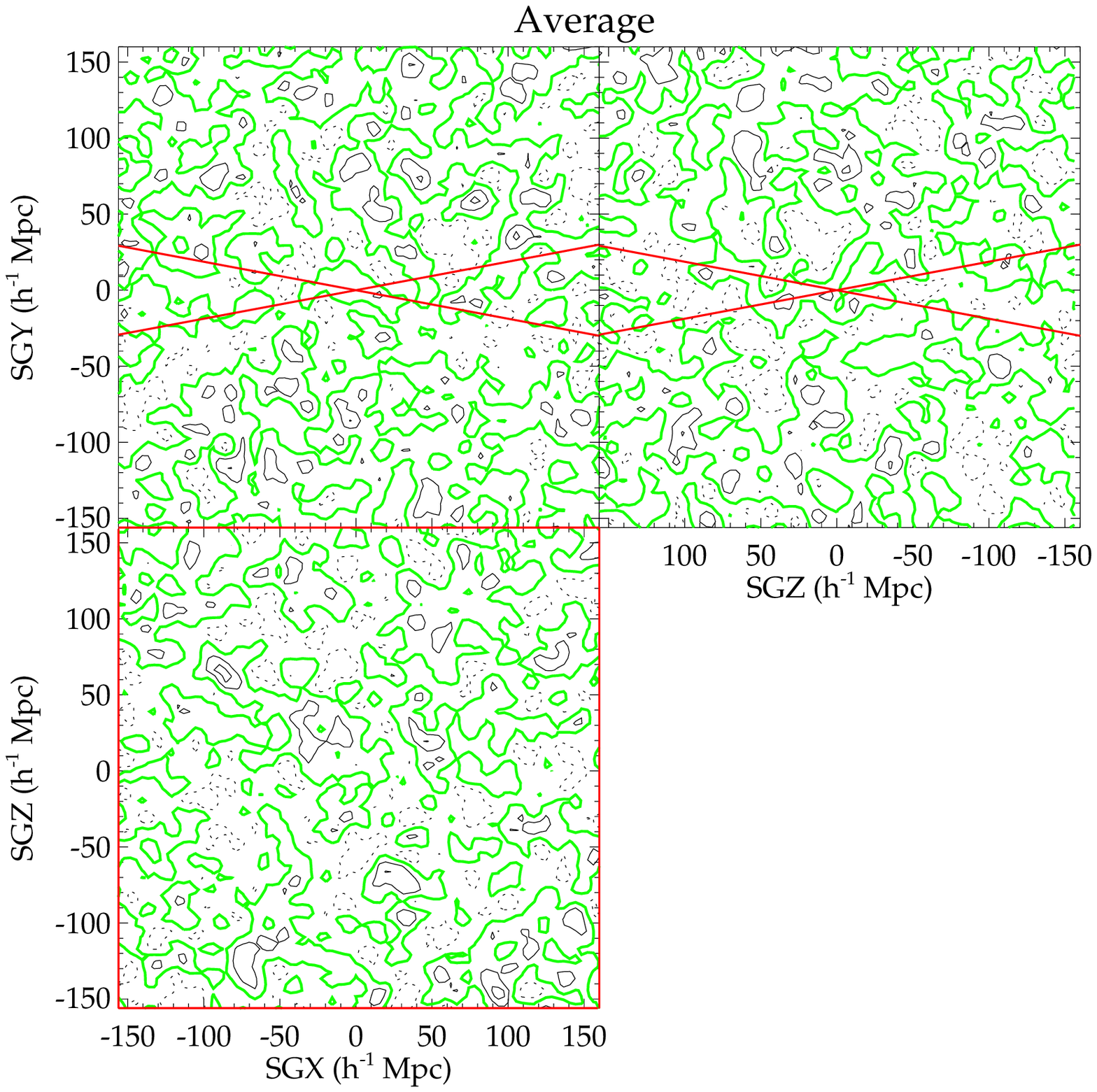}\\ 
\vspace{-0.5cm}
\hspace{-0.58cm} \hspace{-1.5cm} \includegraphics[width=0.55\textwidth]{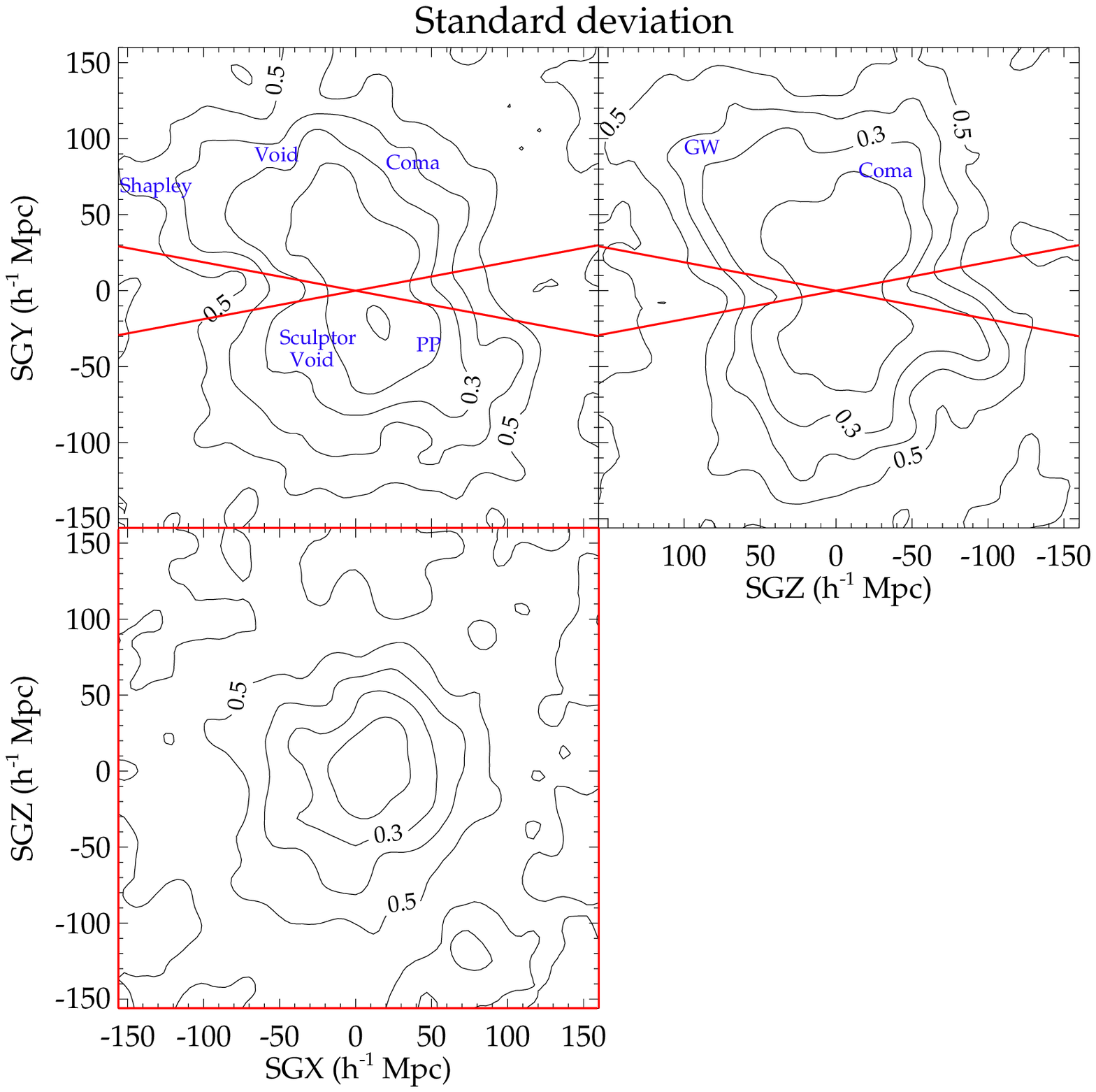}
 \includegraphics[width=0.55\textwidth]{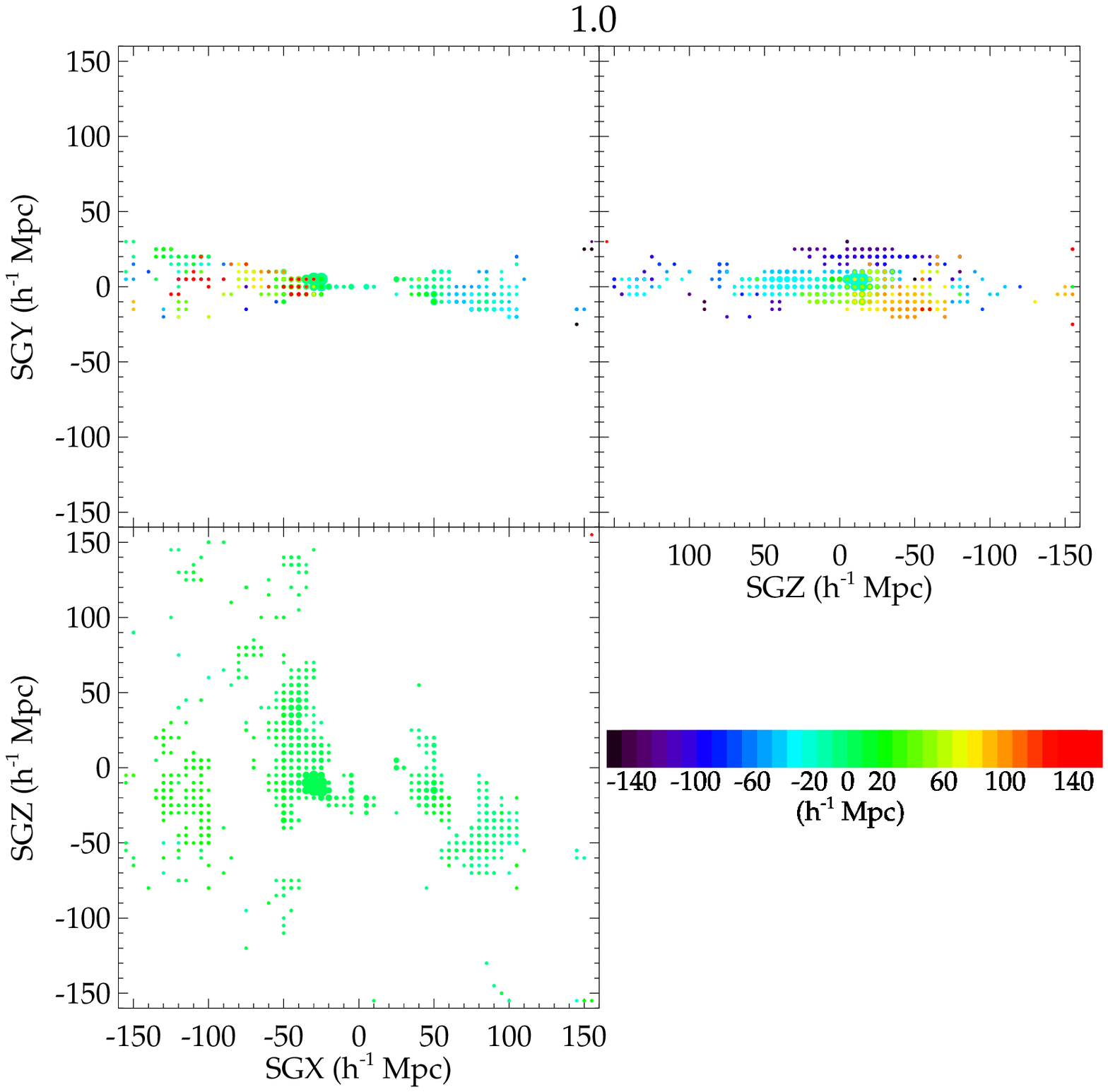}
 \caption{Top panels: Averaged overdensity fields (black solid contours) of 200 constrained (left) and 200 random (right) realization fields in the 2\hMpc\ thick X-Y, Y-Z and X-Z supergalactic slices centered on the observer. The green contours show the transition between overdensities and underdensities ($\delta$ = 0), the dashed contours stand for the underdensity and the red lines delimit the Zone of Avoidance in the different slices,
   corresponding to $(l,b)=(-10^\circ,+10^\circ)$ in Galactic coordinates. The name of a few local structures is given in blue on top of the averaged constrained field (GW stands for `Great Wall' and PP for `Perseus-Pisces'). Bottom panels: Left: Standard deviation of the 200 constrained overdensity fields (black solid contours). Labels give the value of the standard deviation. ; Right: Dots stand for 1.0-overdensity threshold-pass regions with a 3$\sigma$ signal; see the text for a
   more detailed explanation. Dot sizes are proportional to the probability of cells to be threshold-pass. Their location in the third dimension (Z, X and Y directions from left to right, top to
   bottom respectively) is indicated by the color scale.}
\label{fig:space}
\end{figure*}

In the rest of this section, we focus on cells that have a threshold-pass probability with a 3$\sigma$ level of significance. For
instance, for the 1.0-overdensity threshold, the probability that the cell is threshold-pass must be at least 24\% (=15\% + 3$\times$3\%,
i.e.\ the mean of the random probability distribution function plus three times its standard deviation) for the cell to be considered. The
probability distribution being quasi-Gaussian, a positive 3$\sigma$ level is an appropriate choice for selecting the statistically significant cells in the positive tail of the distribution and removing
the others. To locate these cells (or potential structures) in space, the top left panel of Fig.~\ref{fig:space} shows as black contours the
average overdensity field of the 200 CRs in three 2\hMpc\ thick supergalactic slices centered on the observer. Keeping the same contour
values, the same plot is shown for the average overdensity field of the 200 RRs on the right side. The red lines identify the ZOA in the different supergalactic
slices. While the average of the 200 RRs presents a
quasi-null field (green contours) everywhere, there are, as expected from our statistical study of simulations, clearly distinct structures
in the average of the CRs, including in the ZOA that is of interest here. Such an observation means that similar structures are repeatedly at the same
location in the CRs while the locations of similar structures differ substantially from one RR to another. Consequently, while they annihilate each other when averaging in the random case, they reinforce each other in the constrained case. This claim is further supported by the standard deviation of the 200 constrained realizations shown in the bottom left panel of Fig.~\ref{fig:space}: the constrained realization technique decreases the cosmic variance between the different constrained overdensity fields with an increasing effect from the outer to the inner part (where most of the constraints are). However these plots of the slice structures do not show either the entire ZOA nor the likelihood of a structure in the ZOA of the local
Universe. Although the standard deviation indeed highlights the part of the box that is the most constrained, it does not exhibit particular information on each one of the structures especially those in the ZOA.  To get additional information, the $>$3$\sigma$ cells are plotted in the bottom panel of Fig.~\ref{fig:space} with a size proportional to the
probability that they are threshold-pass and a color indicating their distance in the supergalactic direction perpendicular to the plotted slice.

Structures with the highest probabilities (bigger dots) of containing structures are located at nearer distances along the line of sight,
simply because the technique is better able to reconstruct structures when the ZOA is less extended (i.e. nearby). However there are some more
distant structures that also exhibit probabilities with 3$\sigma$ significance. Such structures, further from the regions where 
constraints are (since the ZOA covers a greater physical extent), need to be considered more fully, and compared with observational
expectations and predictions.Such comparisons are conducted in the next section.


\section{Observed and Predicted Structures}

Observers working on the ZOA prefer working in Galactic coordinates (longitude and latitude). To compare the theoretical results
with observations, we convert the grids in supergalactic coordinates to Galactic coordinates, and split the ZOA into small cells of Galactic
longitude, Galactic latitude, and distance, $(l,b,d)$. We compute the overdensities in these cells and gather the cells into partial spherical
shells of depth 10\hMpc. We then determine the overdensity contours on the projected distribution of cells. Several of these contour maps of
overdensity in partial spherical shells, centered on distances between 75 to 215\hMpc, are shown in Figs.~\ref{fig:gl_gb1}, \ref{fig:gl_gb2}, and \ref{fig:gl_gb3}.

Galaxies and clusters observed in the ZOA from various surveys are plotted on top of the contours. colors and symbols are as follows:
(1)~small red filled circles are part of our Nan\c cay observations of the Puppis 3 cluster; 
(2)~small green filled circles are from the HIZOA-S survey \citep{2016AJ....151...52S};
(3)~small blue filled circles are from \citet{2006MNRAS.369.1131R}; 
(4)~small light blue filled circles are from the 2MASS redshift survey \citep{2012ApJS..199...26H} and
(5)~the big red filled circle is a cluster from the CIZA (Clusters In the Zone of Avoidance) project \citep{2002ApJ...580..774E}. Note that
small shifts in the Galactic longitude of galaxy points were applied in two panels (second panel of Fig. \ref{fig:gl_gb2} and second panel of
Fig. \ref{fig:gl_gb3}). These shifts were less than 5$^\circ$. The accuracy of the WF is about 2-3 \hMpc.  A small 5$^\circ$ shift at distances greater  than 50 \hMpc\ is thus completely justified and even expected.

In the rest of this section, the probability quoted for each detected structure is the percentage of CRs in which the structure is detected at the 1.0 overdensity threshold.

\subsection{The Puppis 3 cluster, other galaxy groups and a Cygnus concentration}

\begin{figure}
\vspace{-2.3cm} \hspace{-1cm}
\includegraphics[width=0.6\textwidth]{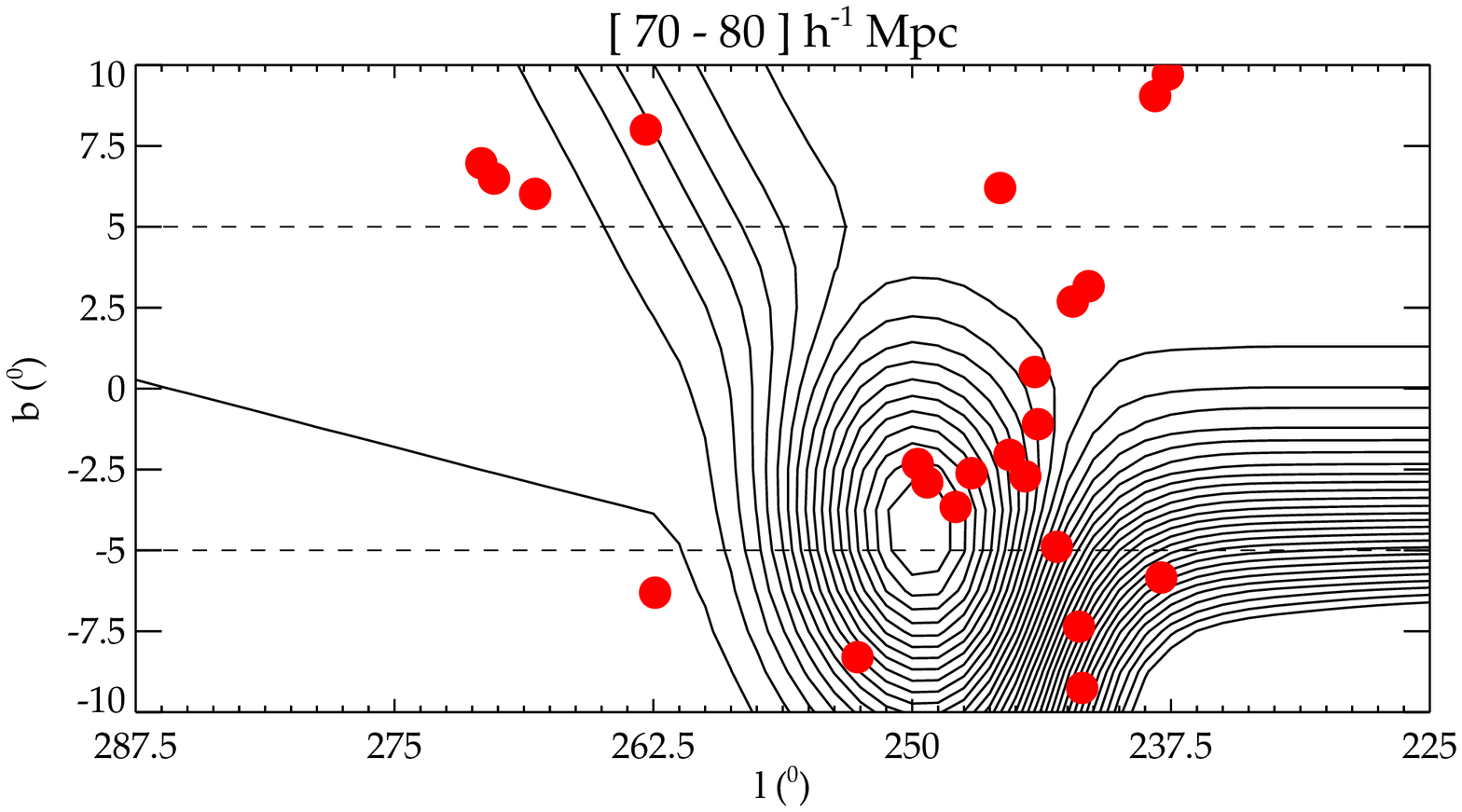}\\
\vspace{-4.2cm}

 \hspace{-1cm}\includegraphics[width=0.6\textwidth]{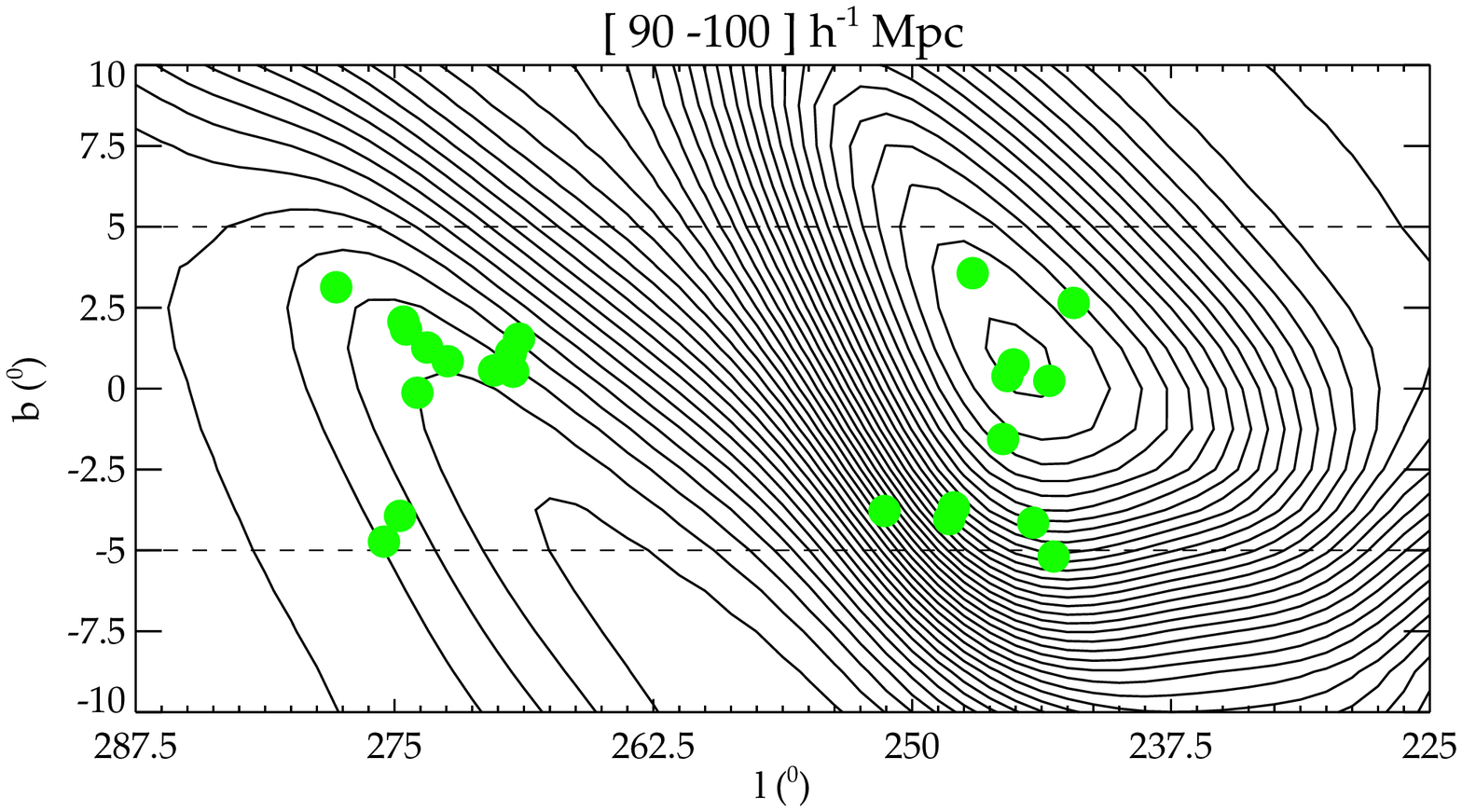}\\
 
\vspace{-4.2cm} \hspace{-1cm}
\includegraphics[width=0.6\textwidth]{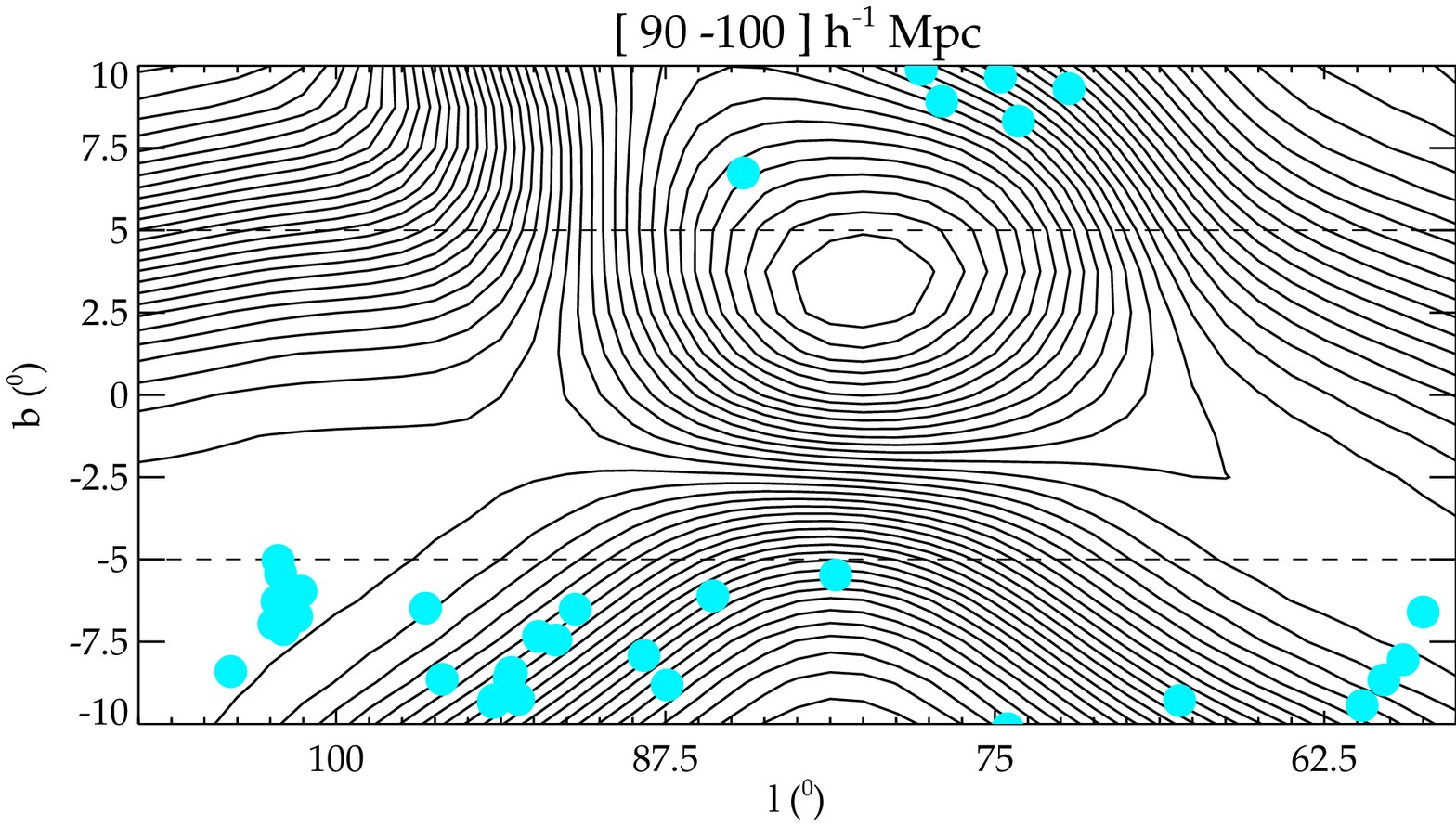}\\
\vspace{-2.cm}
\caption{Galactic coordinate representation of the average overdensity field of 200 constrained realizations (black contours). The partial
  spherical shells shown are 10\hMpc\ thick and are centered on structures in the Zone of Avoidance (ZOA) at distances of 75\hMpc\ and 95\hMpc. Red, green and light blue filled circles are galaxies observed in three different ZOA surveys. There is a remarkable agreement between the  computed overdensity field and the observations of Puppis~3 in the top
  panel. The two bottom panels probe distances of 95\hMpc\ where two galaxy groups have previously been identified (middle panel). The bottom panel suggests the presence of a
  mass concentration in the Cygnus constellation. Galaxies observed on both sides of this concentration reinforce its existence. The observations and the average
  reconstruction are in very good agreement.}
\label{fig:gl_gb1}
\end{figure}

The top panel of Fig.~\ref{fig:gl_gb1} shows a remarkable agreement between the observed galaxies (red filled circles) in the Puppis~3
cluster at about 70\hMpc, first hinted at by \citet{2004MNRAS.347..541C} and confirmed by \citet{2016AJ....151...52S}, and the average 
reconstructed field (black solid contours); the Puppis~3 cluster has a 21.5\% probability in the reconstruction, namely it is significant at the 2$\sigma$ level. Interestingly enough, we note that \citet{2006MNRAS.373...45E} also theoretically predicted that cluster (C11 in their paper). The agreement is remarkable in the sense that they base their reconstruction on the 2MASS redshift survey while we base ours on the second radial peculiar velocity catalog of Cosmicflows, namely the two datasets are of different natures and do not share the same spatial distribution of datapoints.

The middle panel of the figure is remarkable in two ways: (1)~\emph{two} distinct clumps of galaxies at 95\hMpc\ are not only clearly observed but also
reconstructed with about 25\% probabilities (more than a 3$\sigma$ significance level); and (2)~they correspond to galaxy \emph{groups} with lower density than clusters and so would not
necessarily be expected to appear in the reconstruction. These groups were identified by \citet{2007ApJ...655..790C} in the Two Micron All Sky
Survey Extended Source Catalog \citep{2006AJ....131.1163S} using a percolation algorithm. 
Finally, the bottom panel of the figure suggests the presence of a mass concentration in the Cygnus constellation  close to Deneb and the North America Nebula (NGC7000). Galaxies from the 2MASS redshift survey (light blue filled circles) observed on both sides of this concentration strengthen its most probable existence. The structure is significant at more than the 2$\sigma$ level (23\% probability). Actually, \citet{2006MNRAS.373...45E} predicted a structure at about 100 \hMpc, (l,b)=(100$^\circ$,-7$^\circ$) that they call C13, i.e. not far from the aforementioned mass concentration. Looking at this exact location, the probability to have a structure is in our case as high as 28\% (3$\sigma$ significance level).

\subsection{The Cygnus A cluster and two high density peaks}

\begin{figure}
\vspace{-2.3cm}\hspace{-1cm}
\includegraphics[width=0.6\textwidth]{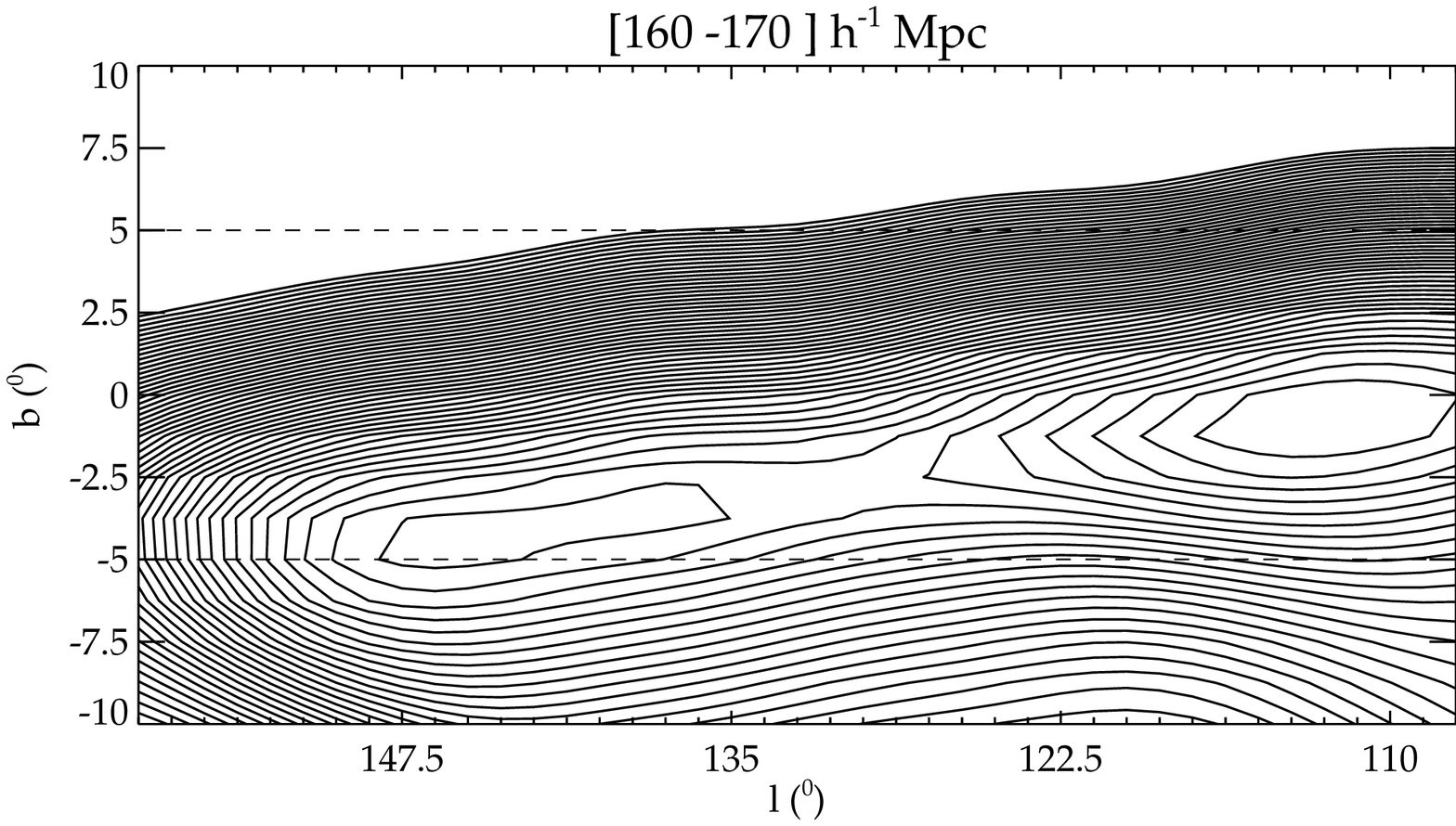}\\
\vspace{-4.2cm}

\hspace{-1cm}\includegraphics[width=0.6\textwidth]{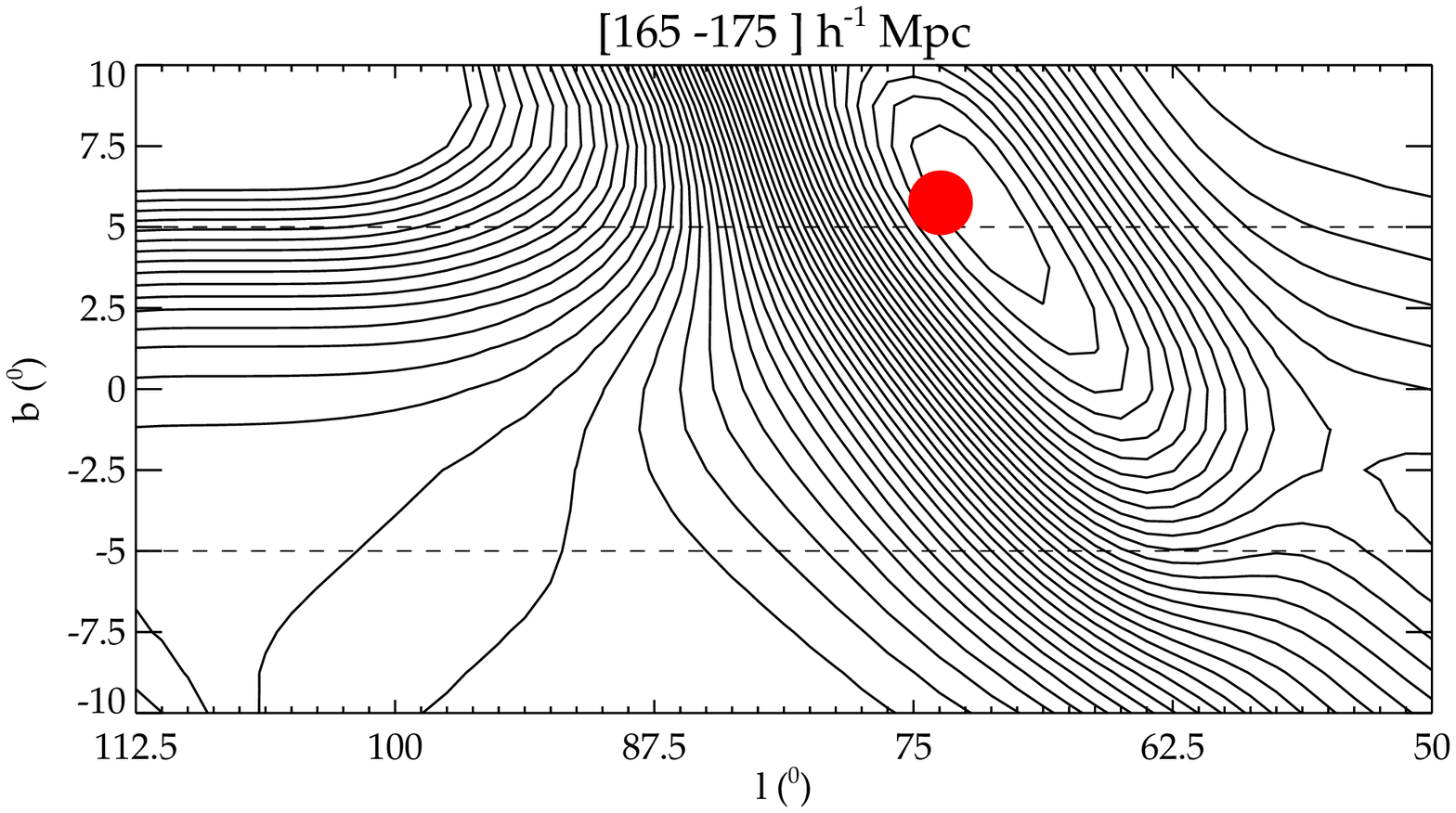}
\vspace{-2.cm}
\caption{Galactic coordinate representation of the average overdensity field of 200 constrained realizations (black contours). The partial
  spherical shells shown are 10\hMpc\ thick and are centered on structures in the Zone of Avoidance (ZOA) around
  $(l,b)=(130^{\circ},0^{\circ})$ at a distance of 165\hMpc\ (top panel) and around $(l,b)=(80^{\circ},0^{\circ})$ at a distance of 170\hMpc\
  (bottom panel). The red filled circle in the bottom panel shows the location of the Cygnus~A cluster observed in the CIZA project at a
  distance about 170\hMpc. There is a remarkable agreement between the computed overdensity field and the location of Cygnus~A. The top panel
  shows two reconstructed structures that need to be observationally confirmed.}
\label{fig:gl_gb2}
\end{figure}

The excellent agreement between the observations and the average reconstructed field at distances less than 100\hMpc\ encourages us to
probe greater distances. Figure~\ref{fig:gl_gb2} shows two slices centered at 165 and 170\hMpc. The top panel shows two structures at
about $(l,b)=(140^{\circ},-4^{\circ})$ and $(l,b)=115^{\circ},-1^{\circ})$ with probabilities of about 22\% and
24\% (respectively 2 and 3$\sigma$ significance level).  Examining the various available online catalogs of galaxies and 
clusters in the ZOA, we were unable to identify particular objects in this region due to a lack of available redshift measurements to confirm
the distances of the objects predicted at these locations. However, here again, there is a good agreement with the predictions from \citet{2006MNRAS.373...45E} although at the edge of the 2MASS redshift survey: they predicted two clusters (C31 and C33 in their paper) at very similar locations as those identified above. This new agreement inspires us to make some deeper comparisons between their predictions and ours although it is not the principal subject of this paper. The results are gathered in Appendix B and reveal astonishingly similar predictions. Combined with the agreement between the X-ray observation 
of the Cygnus~A cluster and its probability of 29.5\% (significance of quasi 5$\sigma$) in the averaged reconstructed field visible in the bottom panel of the same figure, its stimulates further investigations at even larger distances, deeper in the ZOA.

\subsection{The Vela Supercluster}

\begin{figure}
\vspace{-2.3cm}\hspace{-1cm}
\includegraphics[width=0.6\textwidth]{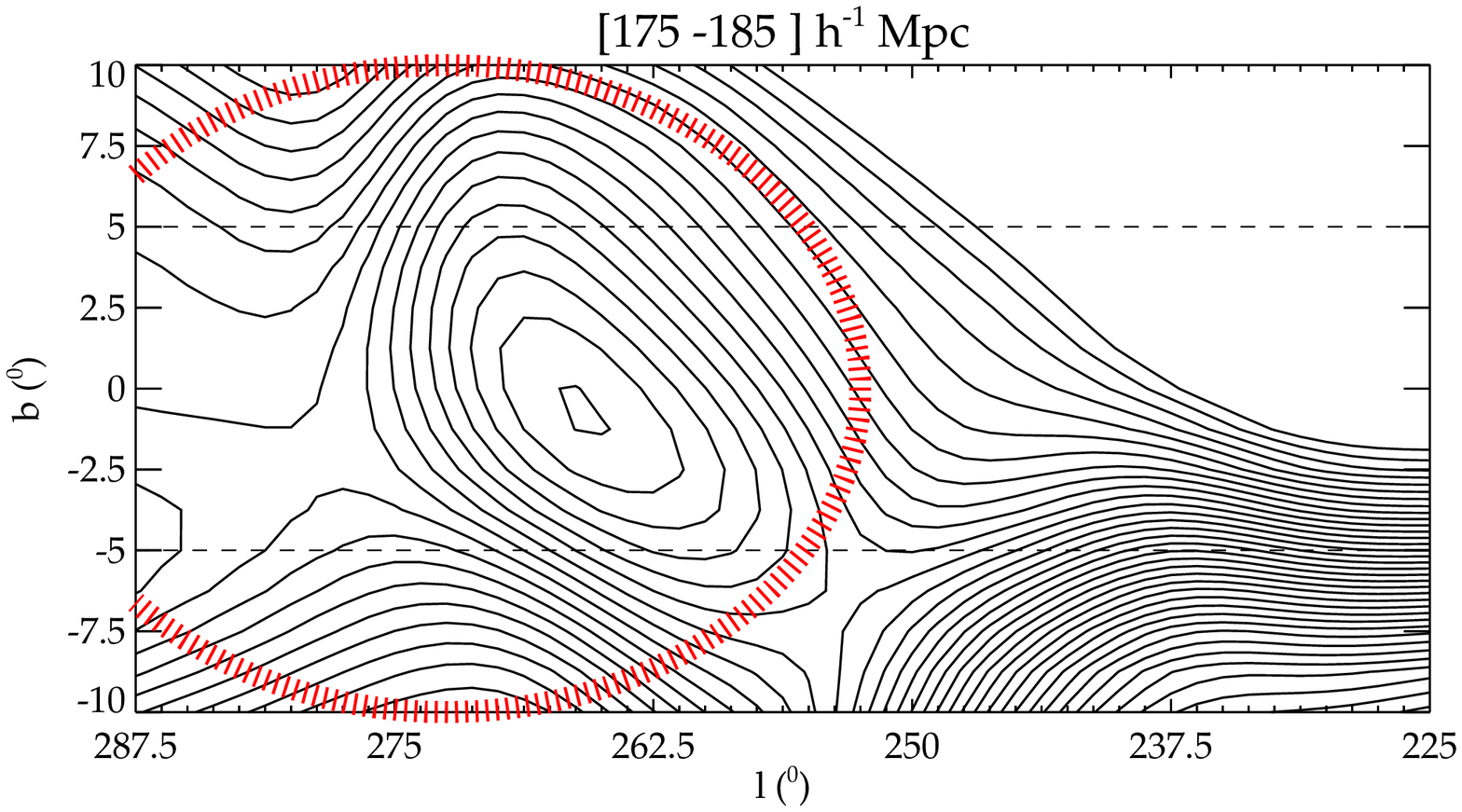}\\
\vspace{-4.2cm}

\hspace{-1cm}\includegraphics[width=0.6\textwidth]{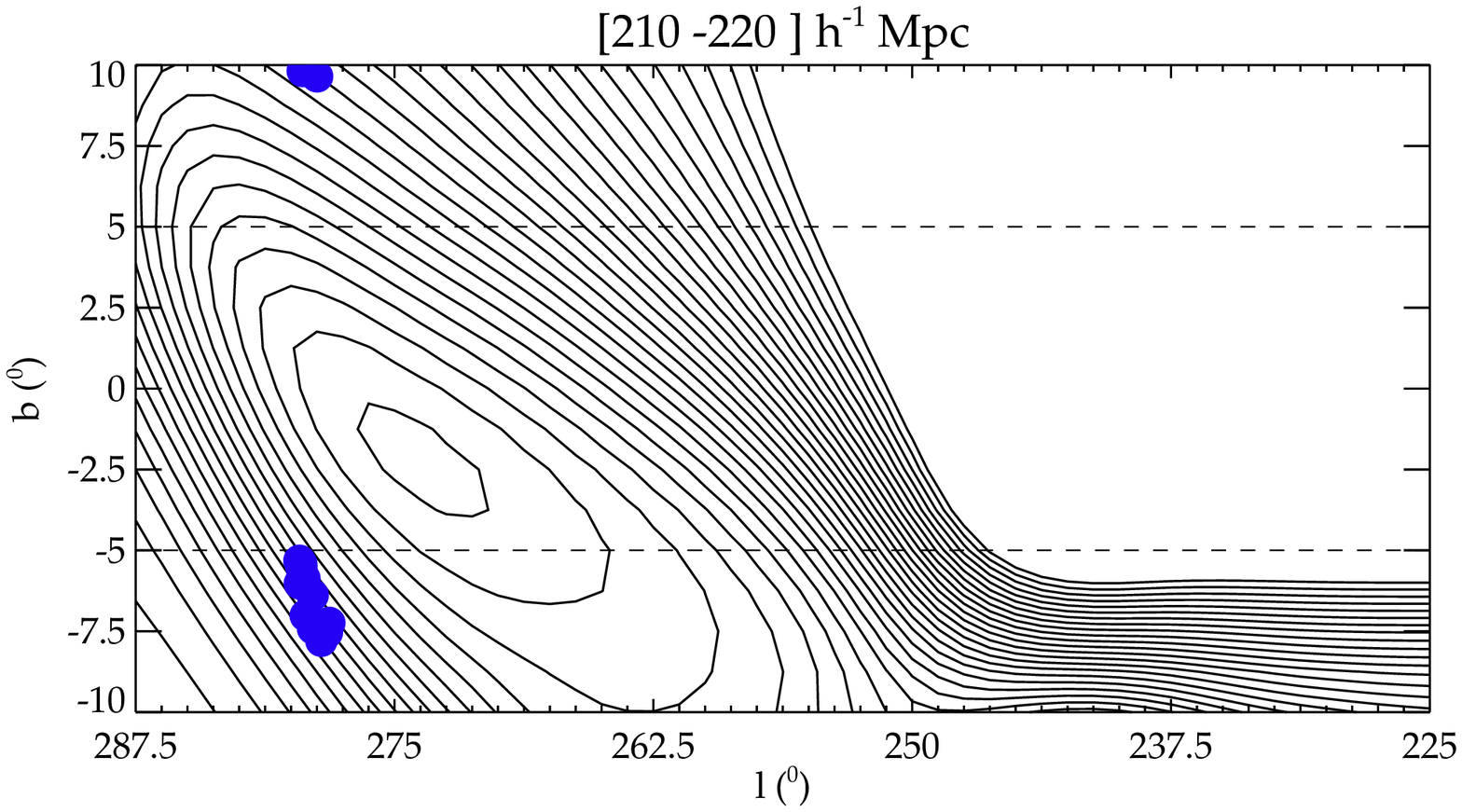}
\vspace{-2.cm}
\caption{Galactic coordinate representation of the average overdensity field of 200 constrained realizations (black contours). The partial
  spherical shells shown are 10\hMpc\ thick and are centered on structures in the Zone of Avoidance (ZOA) around
  $(l,b)=(260^{\circ},0^{\circ})$ at distances of 180\hMpc\ (top panel) and 215\hMpc\ (bottom panel). The red dotted line ellipse in the top
  panel indicates the location of the recently discovered Vela supercluster. There is a remarkable agreement between the prediction
  and the observation. The bottom panel shows the second wall beyond the center of the Vela supercluster.}
\label{fig:gl_gb3}
\end{figure}

Motivated by the recent announcement of the discovery of a supercluster in the Vela region of the ZOA \citep{2017MNRAS.466L..29K}, we plot in
Fig.~\ref{fig:gl_gb3} partial spherical shells at distances of 180 and 215\hMpc\ to probe this region.  In the top panel of the figure, at the
claimed location of the Vela supercluster (denoted by the red dotted line ellipse centered at $(l,b)=(272.5\pm20^{\circ},0\pm10^{\circ})$,
 there is a structure with a 27.5\% probability (i.e.\ 55 of the 200 CRs show the Vela supercluster structure or a 4$\sigma$ significance level). Note the very small
difference between the predicted position of the structure and that announced by \citet{2017MNRAS.466L..29K} based on
observations. \citet{2017MNRAS.466L..29K} describe the region as ``two merging wall-like structures, interspersed with clusters and
groups''. They observed the second wall to be at about 220\kms. Remarkably, in the 10\hMpc\ thick slice centered on 215\hMpc,
this second wall is present with a 23\% probability (almost a 3$\sigma$ significance level). A few galaxies
observed by \citet{2006MNRAS.369.1131R} indeed hint at the extension of a structure deep into the ZOA. Probing intermediate distances between
these two shells, we do not find any prominent structures, reinforcing the claim of two separate walls. All in all, this study reinforces the
claims of \citet{2017MNRAS.466L..29K}: the central core of the Vela supercluster seems to be hidden behind the thickest dust layers of the
Milky Way. If the Vela supercluster is as massive as the Shapley supercluster, it would reduce the difference in both direction and
amplitude between the observed and predicted values of the Local Group dipole motion.

\section{Conclusion}

The ZOA is very often discarded in studies of the local large-scale structure, either because observationally it is too complex and
difficult or because theoretically it is replaced by extrapolated or statistically random structures. However, in the upcoming era of high
precision cosmology, whole-sky coverage is increasingly essential to address issues related to the origin of the CMB dipole, to the largest
structures that possibly exist behind the ZOA, and to the local bulk flows. Consequently, more and more observational programs are taking on
the challenge of observing behind (or through) the Galactic dust. In this paper, we study the ZOA from a theoretical perspective, and
confront observations via the use of constrained realizations of the local Universe. These constrained realizations (CRs) stem from three
components: (1)~local observations of the large-scale structures within 200\hMpc, used as constraints; (2)~a random field to statistically
compensate for missing structures where the data are too sparse and too noisy; and (3)~a prior cosmological model. These constrained
realizations reproduce the linear part of the local overdensity field within 200\hMpc\ very well and, thus, constitute a sufficiently good simulacrum 
of the local Universe for our statistical exploration of structures in the ZOA.

This paper is based on the statistical study of 200 CRs and their 200 random-realization counterparts (RRs; 200 random
realizations sharing the same prior cosmological model and random seed, calculated at a small computational cost). To show that in the random case the structures in the ZOA differ completely from one realization to another, while in the constrained case the structures are distributed in agreement with the surrounding environment, different recipes lead to different probability distribution functions. They are based on the overdensity values of cells of size 5\hMpc\ filling up the entire ZOA region.  These values are compared to a given threshold and probability of each cell
being threshold-pass (i.e.\ having a value above the threshold) is
estimated from the 200 realizations. The probability distribution functions are defined as follows:
\begin{itemize}
\item constrained probability distribution: the procedure described above is looped over the 200 CRs and then reiterated over all the
  cells. The result is the probability distribution of cells in the ZOA being threshold-pass.
\item random probability distribution: same as before but for the 200 RRs.
\item global probability distribution: for a given cell and threshold, once the procedure is looped over the 200 CRs on the one hand and over
  the 200 RRs on the other hand, the two resulting counts are subtracted. The procedure is repeated for all the cells. It gives the
  probability distribution of a cell in the ZOA being threshold-pass once the random probability has been subtracted.
\item individual probability distribution function: for a given cell,
  threshold and random seed, the cell is counted as being threshold-pass  if it meets the requirement of having an overdensity value above the
  threshold only in the CR and not in the RR sharing the same random seed. The procedure is reiterated over the 200 CR/RR pairs and then
  over all the cells in the ZOA. It results in the probability distribution of a cell in the ZOA being threshold-pass solely because
  of the constraints; the random part alone would not have induced an overdensity sufficiently high for the cell to be threshold-pass.
\end{itemize}

If the constraints did not affect the reconstruction of the ZOA, the constrained and random probability distributions would be identical;
this is not the case. While the random probability distribution is a Gaussian, the constrained one is skewed and flatter, 
revealing that cells in the high tail of the constrained distribution most likely host a structure---their probability is increased due to the constraints.

That the random and constrained probability distributions differ is confirmed by the shape of the global probability distribution
function. This function is a Gaussian; this implies that some cells have a probability to host a structure that is greater over 200 constrained realizations than these same 
cells in 200 random realizations associated with the same prior cosmological model. The individual probability distribution is even more
remarkable: it confirms that some cells host structures by the sole action of the constraints; the random part alone would not cause the cells to be threshold-pass.

The probability derived for each cell allows us to locate the reconstructed structures in the ZOA that have the highest probability
(i.e.\ in which we can have the greatest confidence). These structures are confronted with observations. As expected, since the reconstruction
technique works well nearby, where the physical extent of the ZOA is least, there is good overall agreement between reconstructed and
observed structures close by. For instance, the Puppis~3 cluster at about 70\hMpc\ is successfully reconstructed. More notably, this
agreement extends farther away: the Cygnus~A cluster at about 170\hMpc\ is equally well reconstructed. The agreement for these two known
structures confirm the effectiveness of the techniques and encourages us to study even larger distances and deeper into the ZOA. Looking at
180\hMpc, we find a structure reconstructed with 27.5\% probability (i.e. at the 1.0 overdensity threshold there is a structure in 27.5\% of the CRs) that is almost centered on the location of the
recently discovered Vela supercluster \citep{2017MNRAS.466L..29K}. The signal for this structure is significant at the 4$\sigma$ level. This prediction is remarkable in the sense that it is inferred with observational data, 98\% of which are within 160\hMpc\ and 50\% of which are within 61\hMpc. Furthermore, probing slightly larger distances
still, we also see the second wall at about 220\hMpc\ behind the centre of the Vela supercluster also brought to light by \citep{2017MNRAS.466L..29K}; this second wall is reconstructed despite
being outside the region covered by the data used as constraints.

This study lends support to the existence of two merging wall-like structures forming the Vela supercluster, with a central core exactly behind the deepest part of the ZOA. This supercluster might resolve the misalignment of clustering dipoles and account for the residual bulk flows.

\section*{Acknowledgements}

The authors gratefully acknowledge the Gauss Centre for Supercomputing
e.V. (www.gauss-centre.eu) for providing computing time on the GCS
Supercomputer SuperMUC at LRZ Munich. The authors thank the referee for constructive comments that helped to improve the manuscript. This research has made use of the
SIMBAD database and the VizieR catalog access tool, operated at CDS,
Strasbourg, France as well as of the TOPCAT interactive graphical tool
developed by Mark Taylor. JS acknowledges support from the Astronomy
ESFRI and Research Infrastructure Cluster ASTERICS project, funded by
the European Commission under the Horizon 2020 Programme (GA 653477). RKK thanks the South African National Research Foundation for their support.

\appendix\markboth{Appendix}{Appendix}
\renewcommand{\thesection}{\Alph{section}}
\numberwithin{equation}{section}

\begin{appendix}
\section{The Wiener Filter and Constrained Realization techniques}
The Wiener Filter technique is the optimal minimal variance estimator given a dataset and an assumed prior power spectrum. Data dominate the reconstruction in regions where they are dense and accurate. On the opposite when they are noisy and sparse, the reconstruction is a prediction based on the assumed prior model. It schematically multiplies data by $\frac{\mathrm{Power\ Spectrum}}{\mathrm{Power\ Spectrum + Error^2}}$.  Consequently, when data are too noisy or absent, the null field is reached. The Constrained Realization technique (CR) adds the fluctuations of an independent random realization to re-establish statistically structures. In other words, the CR adds a random realization of the residual between the observed field and the minimal variance estimator of this field (the Wiener Filter reconstruction). Briefly, the overdensity $\delta^{CR}$ and velocity \textbf{v$^{CR}$} fields of constrained realizations are expressed in terms of the random realization fields $\delta^{RR}$, \textbf{v$^{RR}$} and the correlation matrixes. For a list of M constraints $c_i$:

\begin{equation}
\delta^{CR}(\textbf{r})=\delta^{RR}(\textbf{r})+\sum_{i=1}^M \langle\delta(\textbf{r})c_i\rangle\eta_i 
\label{eq1}
\end{equation}

\begin{equation}
 v_{\alpha}^{CR}=v_{\alpha}^{RR}(\textbf{r}) + \sum_{i=1}^M \langle v_{\alpha}(\textbf{r})c_i\rangle \eta_i \quad with \quad \alpha=x,y,z
 \label{eq2}
 \end{equation}

where $\eta_i=\sum_{j=1}^M\langle C_i C_j\rangle^{-1}(C_j-\overline{C_j})$ are the components of the correlation vector $\eta$. $\overline C_i$ are random constraints with the noise and $C_i=c_i+\epsilon_i$ are mock or observational constraints plus their uncertainties. Hence, $\langle C_i C_j\rangle$ is equal to $\langle c_i c_j\rangle+\epsilon_i^2\delta_{ij}$ assuming errors constitute a purely statistical noise with a Gaussian distribution. The constraints can be either densities or velocities. $\langle AB \rangle$ notations stand for the correlation functions involving the assumed prior power spectrum.\\
 
The associated correlation functions are given by: 
\[ \langle \delta(\textbf{r}\, ') v_{\alpha} (\textbf{r} \,'+\textbf{r}) \rangle  = \frac{H a f}{(2 \pi)^3}\int_0^\infty \frac{ik_{\alpha}}{k^2}W(kR)P(\textbf{k}) e^{-i\textbf{k}.\textbf{r}}d\textbf{k} \]
\begin{equation} \; = -H a f r_{\alpha} \zeta (r) \end{equation}

\[  \langle v_{\alpha}(\textbf{r} \,')v_{\beta}(\textbf{r}\, '+\textbf{r})\rangle \;\; = \frac{(H a f)^2}{(2\pi)^3}\int_0^\infty \frac{k_{\alpha}k_{\beta}}{k^4}W(kR)P(\textbf{k}) e^{-i \textbf{k} .\textbf{r}} d\textbf{k}  \]
\begin{equation} = (H a f)^2 \Psi_{\alpha\beta} \end{equation}

where P is the assumed prior power spectrum and W is a Gaussian smoothing kernel and R is the smoothing radius.\\

Removing $\delta^{RR}$ and $v_{\alpha}^{RR}$ from equations \ref{eq1} and \ref{eq2} gives the density and velocity fields of the Wiener Filter reconstruction.

\section{Comparisons with the 2MASS redshift survey-based reconstruction of the ZOA}

Reconstructing the local Universe using the 2MASS redshift survey, \citet{2006MNRAS.373...45E} predicted a certain number of structures up to 160 \hMpc. Table \ref{Tbl:1} lists the clusters they predicted in the Zone of Avoidance that have either been observed or that need to be confirmed observationally. The table gives also the probability to have an overdensity at the location of these predicted or observed clusters according to this paper study. The agreement is remarkable. All the clusters but one (C19) are probable at least at the 1$\sigma$ significance level. Most of them are significant at more than 3 or even 4$\sigma$. For obvious reasons, the probabilities are higher the more nearby the cluster is.

\begin{table*}
\begin{center} 
\begin{tabular}{ccrrr@{     }@{       }@{      }@{     }@{       }@{      }@{     }@{       }@{      }@{     }@{       }@{      }@{     }@{       }@{      }@{     }@{       }@{      }@{     }@{       }@{     }@{       }ccrrr}
\hline
\hline
(1) & (2) & (3) & (4) & (5) & (6) & (7) & (8) & (9) & (10) \\
Name & distance (\hMpc) & l ($^\circ$)& b($^\circ$) & Probability (\%)& Name & distance (\hMpc) & l ($^\circ$)& b($^\circ$) & Probability (\%)\\
\hline
Puppis &	20 & 240 &	-5	&56 & C20	&120	&160	&-10	&35.5\\
C1&	20 &	200	&5	&59.5 & C19	&120	&120	&-10	&27.5\\
C2&	40&	345	&10&	48 & C18	&120	&100&	3	&28.5\\
C8&	40&	295&	7	&76.5 & C17	&120&	15&	-7&	25\\
C5&	40&	195&	0	&24.5 & C21&	120&	190&	7	&28.5\\
Peg&	40&	95	&10	&35 & C19&	140&	115&	-7&	20.5\\
C$\beta$&	40	&120	&7	&34 & C25&	140&	75	&-7	&27\\
Or&	60	&165&	5	&42.5 & C29	&140&330	&-10&	28.5\\
C9&	60	&240	&8&	23 & C30	&140	&300	&10	&30.5\\
Hyd&60	&285	&5	&48 & C21&	140&	195&	-3&	26\\
C8	&60	&310	&7&	36 & C31&  160&	160&	-10	&25\\
A3627&	60	&330&	-5	&30.5 & C33&  160&	110&	0	&23\\
C12&	80	&330&	-7	&26 & C32&	160&	45	&0	&28.5\\
C11 - Puppis3	&80&	250&	-4	&21.5 & C30&	160	&290	&5	&26\\
C13	&100	&100	&-7&	28\\
\hline
\hline
\end{tabular}
\end{center}
\vspace{-0.25cm}
\caption{Comparisons between clusters predicted by \citet{2006MNRAS.373...45E} and this paper: (1) \& (6) Name of the cluster as given by \citet{2006MNRAS.373...45E} or from observations, (2) \& (7) distance of the cluster in \hMpc, (3),(4) \& (8),(9) galactic coordinates in degrees and (5) \& (10) probability of the cluster according to this paper study.}
\label{Tbl:1}
\end{table*}

\end{appendix}

\bibliographystyle{mnras}

\bibliography{biblicomplete}

\label{lastpage}

\end{document}